\documentclass{aa}  

\usepackage{hyperref}
\usepackage{graphicx}
\usepackage{txfonts}
\newcommand{\xmm}{{\em XMM-Newton}}

\newcommand{\rhoOph}{$\rho$~Oph}
\newcommand{\pn}{{\em pn}}
\newcommand{\fxu}{{erg~s$^{-1}$~cm$^{-2}$}}
\newcommand{\lxu}{{erg~s$^{-1}$}}
\begin{document}

   \title{The early B-type star Rho Oph A is an X-ray lighthouse}

   \author{Ignazio Pillitteri
          \inst{1,2}
          \and
          Scott J. Wolk\inst{2}
          \and
          Fabio Reale\inst{1}
          \and
          Lida Oskinova\inst{3}
          }

   \institute{INAF-Ossevatorio Astronomico Palermo, Piazza del Parlamento 1, 90134 Palermo Italy\\
              \email{pilli@astropa.inaf.it}
         \and
             Harvard-Smithsonian CfA, 60 Garden st Cambridge MA 02138
          \and
             Institut f\"ur Physik und Astronomie, Universit\"at Potsdam
              Karl-Liebknecht-Strasse 24/25 14476 Potsdam-Golm, Germany
             }

   \date{Received; accepted }

 
  \abstract
{We present the results of a 140 ks \xmm\ observation of the B2 star \rhoOph~A. The star
has exhibited strong X-ray variability: a cusp-shaped increase of rate, similar to that which we partially observed in 2013, and a bright flare. 
These events are separated in time by about 104 ks,  
which likely corresponds to the rotational period of the star (1.2 days). 
Time resolved spectroscopy of the X-ray spectra shows that the first event 
is { caused by} an increase of the plasma emission measure,
while the second increase of rate is a major flare with temperatures in excess of 60 MK 
($kT\sim5$ keV). From the analysis of its rise, we infer a magnetic field  of $\ge300$ G and a 
size of the flaring region of $\sim1.4-1.9\times10^{11}$ cm, which corresponds 
to $\sim25\%-30\%$  of the stellar radius.
We speculate that either an intrinsic magnetism that produces a hot spot on its
surface or an unknown low mass companion are the source of such X-rays and variability. 
A hot spot of magnetic origin should be a stable structure 
over a time span of $\ge$2.5 years, and suggests an overall large scale dipolar magnetic field
that produces an extended feature on the stellar surface.
In the second scenario, a low mass unknown companion is the emitter of X-rays and it should orbit 
extremely close to the surface of the primary in a locked spin-orbit configuration, 
almost on the verge of collapsing onto the primary. 
As such, the X-ray activity of the secondary star would be enhanced by its young age, and the tight orbit 
 as in RS Cvn systems and \rhoOph\ would constitute an extreme system that is worthy of further investigation. }

   \keywords{stars: activity -- stars: individual (Rho-Ophiuchi) -- stars: early-type 
          -- stars: magnetic field -- stars: starspots -- X-rays: stars}

   \maketitle
%

\section{Introduction}
The emission of X-rays from  massive stars is explained as due to strong stellar winds 
and shocks in O  stars through a mechanism of line deshadowing instability 
(LDI; \citealp{Feldmeier1997a,Feldmeier1997b,Owocki1998}). 
In the case of binarity, wind-wind collision gives an additional source of X-rays, 
sometimes coupled with the presence of magnetic fields
that drive ionized winds and increase shock temperatures \citep{Babel1997,ud-Doula2002}.
The production of X-rays by stellar winds appears less realistic in B-type stars because of
their weaker winds with respect to those from O and WR stars. 

Moving from O to early B-type stars, the rate of detection in X-rays among B stars
falls to about 50\%, where hard X-ray emission in a few cases are a signature of the 
presence of strong magnetic fields or due to an unknown, low mass young and active companion. 
X-rays from single early B stars are observed in a few cases, and their origin is likely 
linked to their strong magnetism. 
Cases of spots in young stars of NGC~2264, which presumably have a magnetic origin, 
are given by \citet{Fossati2014}. Recent spectroscopic surveys of O-B stars have revealed
that about 7\% of these stars are magnetic \citep{Wade2014,Fossati2015}.
In a few cases magnetic fields of strength of a few kG are measured, likely accompanied
by peculiar photospheric abundances.   

\rhoOph\ is a multiple system of B-type stars. They are part of one of the closest 
and densest sites of active star formation. In particular
\rhoOph~A+B is a binary system of two B2 stars separated by about 310 AU at a distance
of $111\pm10$ pc from the Sun (parallax $9\pm0.9$ mas, 
separation $\sim2.8\arcsec$; \citealp{VanLeeuwen2007,Malkov2012}), 
the orbital period of the system is about $2400\pm330$ years \citep{Malkov2012}. 
\rhoOph~A rotates with $v\sim300$ km/s \citep{VanBelle2012,Glebocki2005,Uesugi1982}, 
has a mass of about 8-9 M$_\odot$, and radius of $\sim8 \mathrm{R}_\odot$.

In 2013 we observed \rhoOph\ with \xmm\ discovering that it emits X-rays. 
In particular, we observed a significant rise of X-ray flux in the last 20-25 ks 
of the exposure, accompanied by plasma  temperatures around $\sim3$ keV 
\citep[][hereafter Paper I]{Pillitteri2014c}. 
We hypothesized that either an active spot was emerging on the stellar surface, 
or an unknown low mass companion was causing the feature. 
Along with \rhoOph, we discovered a young cluster of about 25 pre-main 
sequence (PMS) stars, which { are} mostly without circumstellar disks and of $5-10$ Myr in age. 
These stars were likely born together with the \rhoOph\ stars during the first event 
of star formation in the cloud \citep{Pillitteri2016}. 

In this paper we present the results of a follow-up observation of \rhoOph\ 
with \xmm, with a duration of 140 ks. The aim of the observation was to monitor the X-ray 
emission of \rhoOph\ for a full rotational period of the star ($\sim1.2$ days or 104 ks) 
and understand the origin of its X-rays. 
The structure of the paper is the following: in Sect. \ref{observations} we describe the 
observations and data analysis; Sect. \ref{results} present the results; 
and in Sect. \ref{discussion} we discuss them and  present our
conclusions. 


\begin{figure}
\resizebox{\columnwidth}{!}
          {\includegraphics{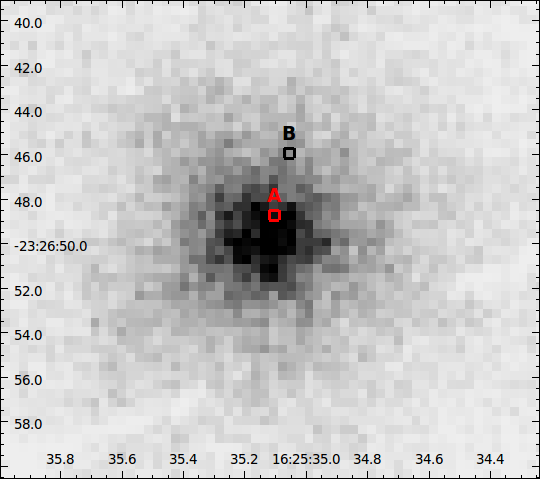}}
\caption{\label{zoom} MOS 1 image of \rhoOph, the positions of the two components from
SIMBAD database are indicated with colored boxes. The centroid of X-ray events is much closer 
to the A component. The scale is linear, the events have been rebinned in blocks of 0.4\arcsec
and smoothed with a Gaussian of $\sigma = 1$ pixel. The separation between the two components
is $\sim2.8\arcsec$. The  core of the point spread function of MOS1 is about 2\arcsec and 
the astrometric precision is $\sim1.2\arcsec$.}
\end{figure}

\section{Observations and data analysis} \label{observations}
\xmm\ observed \rhoOph\ on February 22 2016 for 140 ks (ObsId 0760900101).
We used EPIC camera as the first instrument with the {\it Thick} filter to prevent UV 
leakage due to the brightness of the target (U$=4.30$). For the same reason, OM exposures
were not taken for the safety of the instrument.
The observation data files (ODFs) were reduced with SAS ver. 15.0 to obtain tables of events
for MOS 1, 2 and \pn\ filtered in 0.3-8.0 keV, 
and with filters $\mathrm{FLAG}=0$ and $\mathrm{PATTERN}<=12$
as prescribed by the SAS guide. We used a circular
region of radius 35$\arcsec$ for both source and background events to select the events of \rhoOph. 
For \pn\ we used a background extracted from a region at the same distance from the readout node of 
the same chip of the source, as prescribed by the SAS guide book.  
In Paper I we associated the X-ray emission with the A component of \rhoOph; the MOS1 image 
in Fig. \ref{zoom} shows that this is a reasonable assumption, given that the 
centroid of the X-ray events { is} much closer to the position of \rhoOph~A and completely offset 
with respect to the position of \rhoOph~B.
\begin{figure*}
\resizebox{\textwidth}{!}{
                \includegraphics{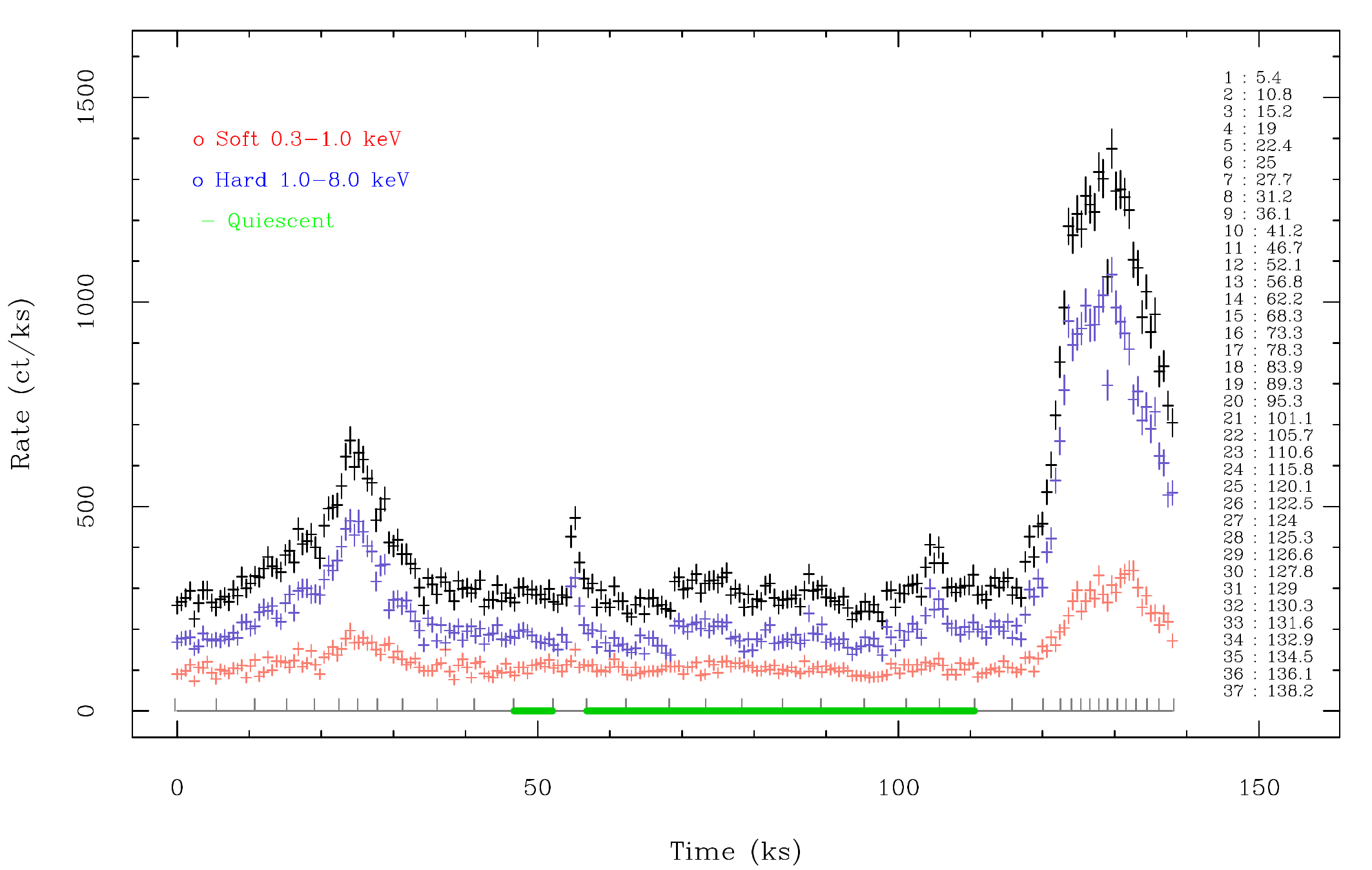}
                }
\caption{ Plot of\label{lcpn} \pn\ light curve of \rhoOph: full band ($0.3-8.0$ keV, black),
soft band ($0.3-1.0$ keV, red), and hard band ($1.0-8.0$ keV, blue). 
Segments used in time resolved spectroscopy are indicated by ticks on the bottom scale and
listed in the column to the right.}
\end{figure*}

Spectra, response matrices (rmf), ancillary files (arf), and light curves of \rhoOph~A 
were obtained with the specific SAS tasks ({\em evselect}, {\em rmfgen}, {\em arfgen}, {\em backscale}, and
{\em specgroup}).  The spectra were analyzed with XSPEC ver. 12.8.0, while generic software R language 
scripts and custom plotting routines were used to produce plots and calculate derived 
quantities of interest. 

To understand the physical changes in the emitting plasma of \rhoOph~A, we divided
{ the \pn\ } light curve in intervals and accumulated the spectra from each time interval.
We chose to obtain spectra with about 1600 counts each, which represents a trade off
to preserve details of the temporal changes of the spectrum and its count statistics.
In this way we divided the light curve in 37 time intervals of variable duration 
($\sim1.2-6.0$ ks, see Fig. \ref{lcpn}). For each time bin we obtained \pn\ spectra
and related calibration files (rmf and arf).
We used a thermal model with absorption composed of two {\em APEC} components 
and a global {\em PHABS} photoelectric absorption ($N_H$). 
We kept $N_H$ fixed to $3\times10^{21}$ cm$^{-2}$ (see Paper I), formally $Z = 0.3 Z_\odot$, 
while temperatures and normalization factors were left free to vary. 
Although the metallicity of Galactic B stars is well established \citep{Nieva2012}, { sub-solar}
abundances reflect a peculiar behavior observed in active stellar coronae (see Sect. \ref{timeresspec}).
\xmm\ allows us to acquire simultaneously high resolution spectra of the central target 
with RGS gratings, provided it is bright enough. 
This was the case for \rhoOph\ in the present 140 ks exposure. 
The RGS spectra were obtained by first extracting the RGS1 and RGS2 spectra of \rhoOph,
then we added together the first order spectra with the SAS task {\em rgscombine}.
We accumulated RGS 
spectra in several time intervals: full exposure (140 ks), first event ($10-40$ ks since 
start of exposure), second event ($115-140$ ks), the sum of first and second
event (high state spectrum),  and the relatively low activity interval in between ($40-115$ ks). 
For each time window we added together both RGS1 and RGS2 spectra of the first order.
We focused on the range $5-20$ \AA\ because this is the range over which 
most of the flux and its changes are observed. 

\rhoOph\ has been observed with ESO-VLT and UVES spectrograph in 2001 and 2005
with different coverage of wavelengths (from 3000\AA\ to 9000\AA), different exposure times 
(2s. to 15s.), and { with a spectral} resolution $R\sim 42,000$. 
We used the reduced spectra from such observations to derive an estimate of the rotational 
velocity along the line-of-sight v~sin$i$ and to check the presence of lines from a low mass companion and any
Doppler shift that could hint at the presence of such companion. 
We used a combination of {\em Iraf/pyraf} and a set of custom scripts to read the spectra 
in Iraf from the original {\em MIDAS} format, display the spectra, measure line absorption, 
remove the heliocentric component of Doppler shift due to the motion of the Earth, and
calculate the cross-correlation function between spectra at two different epochs. 

\section{Results} \label{results}
Fig. \ref{lcpn} shows the light curve of \rhoOph~A in the full band ($0.3-8.0$ keV), soft band 
($0.3-1.0$ keV) and  hard band ($1.0-8.0$ keV). During the observation
\rhoOph~A exhibited two main episodes of variability: one at the beginning of the observation and
another more powerful toward the end of the exposure. 
The first episode was a cusp-shaped increase of rate occurring during the first
40 ks. The peak of this event occurred at $t\sim$ 25 ks and the decay phase appeared to be 
slightly steeper  than the rise phase. 

The second large increase of the rate occurred at $t\sim116$ ks with a peak at around $t\sim129$ ks 
and a decay that lasted until the end of the observation ($t\sim139$ ks). Likely, we missed the very
end of this decay. 
Minor episodes of variability in the form of flares are visible at $t\sim55$ ks and $105$ ks. 
However, in the following we refer to this interval as the quiescent state, while
the main two events of variability are referred to as the high state of \rhoOph. 
The characteristics of the plasma and its changes in temperature and emission
measure are discussed in detail in Sect. \ref{timeresspec}.

\subsection{Phased light curves and rotational velocity}
\begin{figure}
\resizebox{\columnwidth}{!}{
                \includegraphics{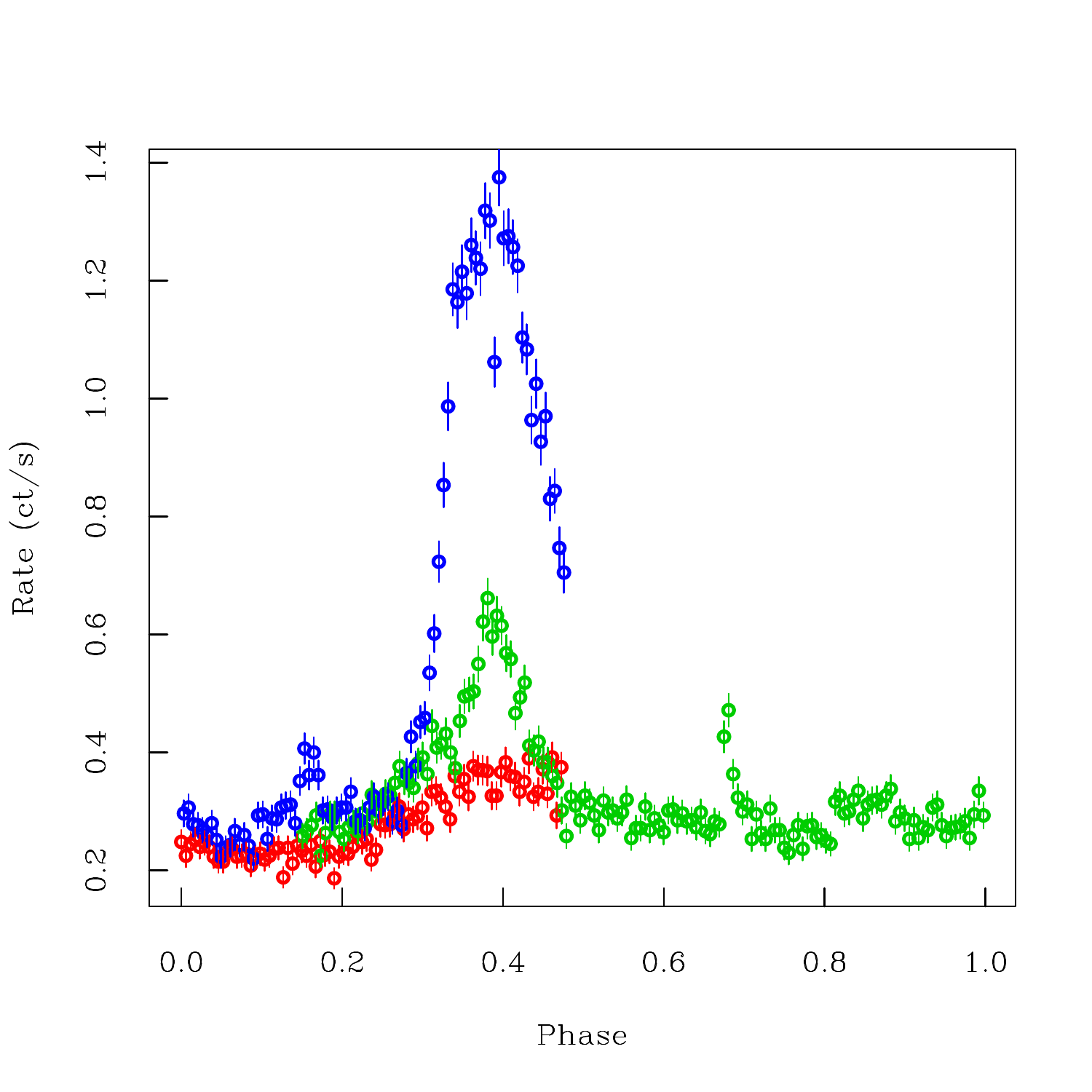}
                }
\caption{\label{phasedlc} Light curves of the rate phase-folded with a period of 1.205 days.
The red dots indicate the 2013 observation, and the green and blue dots indicate the 2016 observation split in two
periods. Phase zero was set arbitrarily to the beginning of 2013 observation. 
The three main peaks of X-ray emission can be put in phase with a period of 1.205 days, 
which corresponds to the rotational period of the star.}
\end{figure}

The rise of the rate in the first event is suspiciously  
similar to the rise of the rate observed in 2013 (Paper I), so we wonder whether the same mechanism 
is responsible for the increase of the rate observed twice in 2016. In particular, 
we hypothesize that a spot or an unseen companion transited during the observation, 
as speculated in Paper I. 
The time elapsed between the two main peaks of the light curve in Fig. \ref{lcpn} is 
approximately 104 ks (or 1.2 days). 
If we consider this time as an initial guess for a phase-folding time, 
only a small adjustment of the period is needed to aligning the rate peaks observed in 2013 and 2016, 
respectively. Fig. \ref{phasedlc} shows the phase folded light curves of 2013 and 2016, where
as zero phase we used the beginning of the 2013 \pn\ light curve. We thus find that a period of 104.11 ks 
corresponding to 1.205 days aligns the three peaks, and this is our best estimate of the period
of rotation of \rhoOph~A based on the periodic variations of its X-ray emission. 
This period corresponds to a rotational velocity of $\sim340$ km/s at the equator when assuming 
a stellar radius of $\sim8 R_\odot$; the { velocity derived from the 1.205 days period } 
is roughly consistent with the rotational velocity of 
\rhoOph~A determined from optical observations \citep[see][]{VanBelle2012}.
The separation between the two epochs of observations is on the order of 1000 days and this 
leads to a precision of the period of 1/1000 of day. 
Additional X-ray observations with shorter cadence could validate the measurement of 
the X-ray rotational period of \rhoOph~A.

\begin{figure}
\resizebox{\columnwidth}{!}{
                \includegraphics{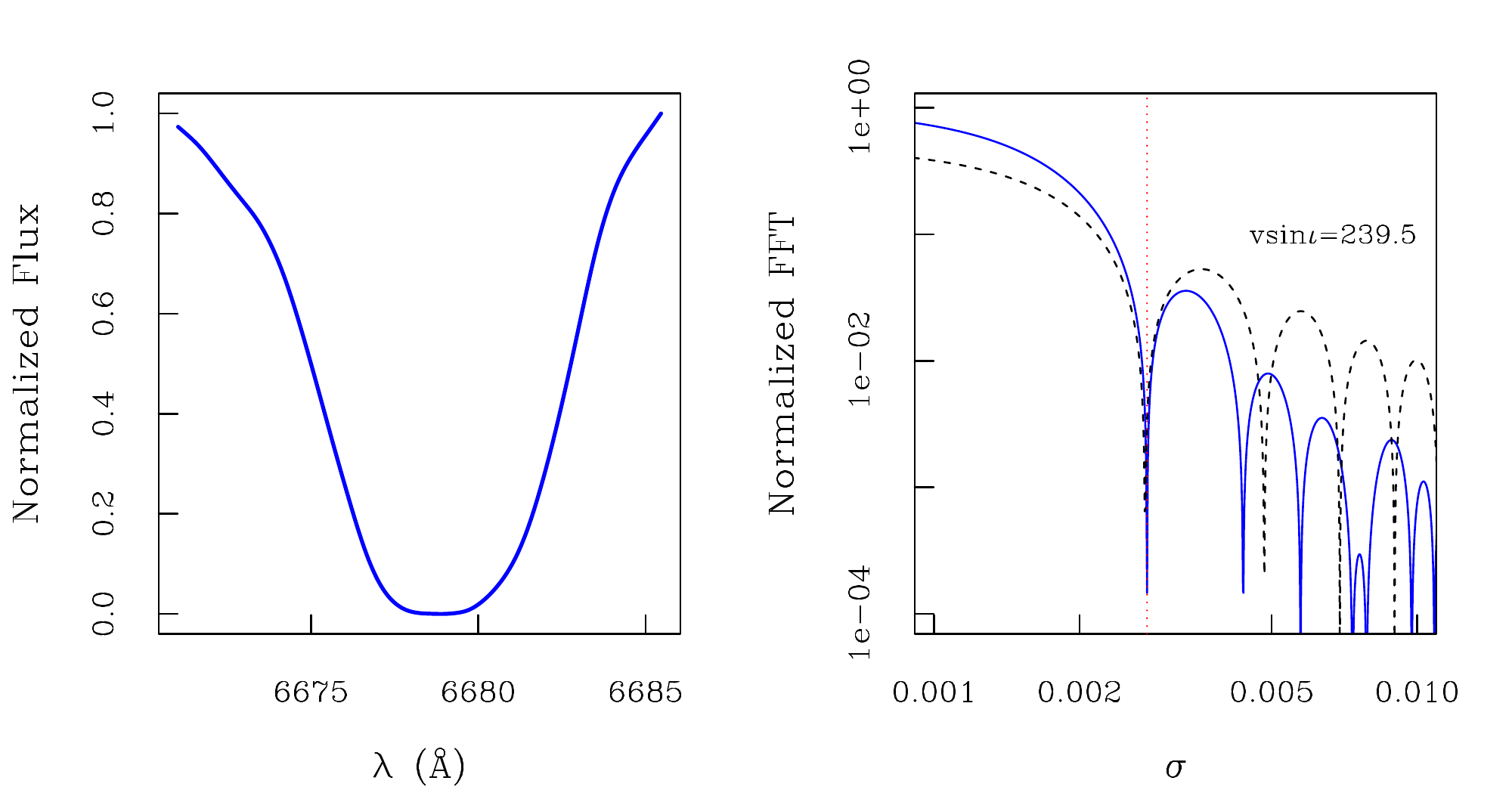}
                }
\caption{\label{fft} Smoothed normalized profile of He line at 6678\AA\ (left) and normalized
amplitude of Fourier transform of the same profile (right, solid line). The dotted curve denotes
the transform of the rotational broadening profile alone. 
The first minimum occurs at $\sigma\sim0.00275\ c/\mathrm\AA$ (vertical dotted line), 
corresponding to v~sin$i\sim 239.5$ km/s. Uncertainty on v~sin$i$ is estimated at around $\pm10$ km/s.}
\end{figure}

We refined the estimate of v$\sin i$ of \rhoOph~A with an analysis of the 
Fourier transform of the line profile of He line at 6678\AA\ from one of the available UVES 
spectra 
(Fig. \ref{fft}, \citealp{Gray88,Smith1976}). The first minimum of the transform is related to 
the line broadening due to rotation and thus to v~sin$i$.
The He line was chosen because it is well isolated from nearby lines, has a good signal, 
and offers an easy modeling of its profile. 
We used a window of wavelengths that encompasses the line (6671\AA$-$6685.5\AA), 
and we smoothed and normalized the profile with a {\em lowess}\footnote{As implemented in R,
\url{https://stat.ethz.ch/R-manual/R-devel/library/stats/html/lowess.html}} smoother.
The Fourier transform was obtained on a sample of 5000 points interpolated along 
the line profile. The first minimum of the Fourier transform occurs at 
$\sigma = 0.00275$c/\AA , which corresponds to v~sin$i=239.5$ km/s. 
The uncertainty is estimated at $\pm10$ km/s.
From v~sin$i$ and the rotational velocity, derived from the period obtained from the X-ray 
variability, we infer a $i$ angle between line of sight and rotational axis of $45\pm5\degr$, 
taking into account an uncertainty of $0.5R_\odot$ on the stellar radius.
As first proposed in Paper I and discussed further in the next section, the presence of an active spot of
magnetic origin and its periodic appearance would imply a magnetic field misaligned with 
the rotation axis. 

On the other hand, if the X-rays were entirely produced by an active low mass companion, depending
on the mass ratio, its motion around \rhoOph\ at a short distance would produce a wobble 
detectable as a Doppler shift in the spectra. For example, supposing a mass ratio of 1:15 
(a late-K star), a period of 1.2 days,  and an orbital velocity of the companion of 300-350 km/s 
would produce a Doppler shift of $\sim20$ km/s, which is easily detectable in the spectra of \rhoOph. 
We first { subtracted} the component due to the motion of the Earth from
the UVES spectra, { then we cross-correlated the spectra taken at different epochs and of similar wavelength} 
coverage \citep{Tonry1979} avoiding regions with telluric lines. We did not detect any Doppler shift, 
meaning that either there is no companion or that  the spectra  
were taken at approximately the same phase ($\mathrm{HJD_1}=2452132.6579; \mathrm{HJD_2}=2453618.4866$, 
$\Delta \phi\sim 0.053$, for a period of 1.205 days). 
The phase difference implies a displacement of $19\degr$ on the orbit. By supposing that \rhoOph~A
has an average $v_{orb}\sim 20 km/s$, the sin$i$ factor would amount to a difference of velocity of 
6.5 km/s at the two phases; this difference is very similar to the UVES spectral resolution. 
A dedicated spectroscopic monitoring is thus required to detect a companion.


\subsection{Time resolved spectroscopy}
\label{timeresspec}
Here we discuss the spectral variations of \rhoOph~A with particular regard to the 
two main episodes of variability that characterized the X-ray light curve of
\rhoOph~A. 
For each time interval, we performed a best fit of the spectrum with an absorbed thermal model composed
of the sum of two APEC models. The free parameters were the temperatures and the respective emission
measures of the thermal components, while absorption and abundances were kept fixed at 
$N_H=3\times10^{21}$ cm$^{-2}$ and $Z=0.3 Z_\odot$ in agreement with what was found in Paper I.  
By letting absorption vary, we obtained values of $N_H$ consistent with the assumed value,
and the variation of plasma temperatures were minimal with respect to the best fit with 
a fixed absorption.
We discuss in detail the evolution of the hot component of 
the plasma, since the cool component had little variation during the observation, 
being comprised of { 0.7 keV $< kT <$ 1.2 keV} with a median of { $kT \sim 0.9$} keV
and standard deviation of 0.15 keV.

\begin{figure}
\resizebox{\columnwidth}{!}{
                            \includegraphics{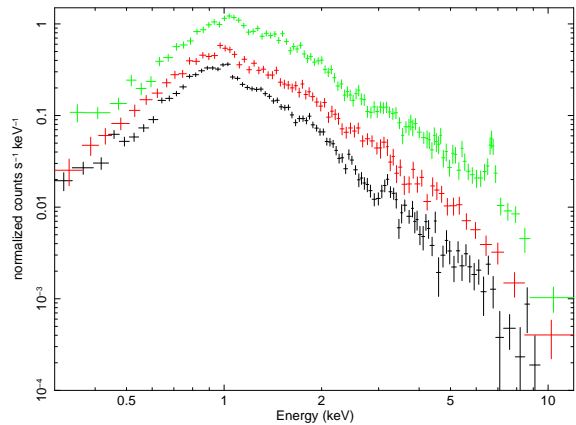}
                           }
\caption{Plot of\label{epicspectra} \pn\ spectra of \rhoOph~A during quiescent phase (time bins $10-24$, 
bin 13 excluded, black spectrum), peak of the first event (bins $5-8$, red spectrum), 
and peak of the second event (bins $27-33$, green spectrum). This latter shows the complex { of} lines of 
ionized Fe at 6.7 keV evidencing the high temperature of the plasma ($\ge5.4$ keV) reached 
during the peak of the second event. }
\end{figure}

 In Fig. \ref{epicspectra} we show the \pn\ spectra of the quiescent phase at the peak of the two
main events of variability. The spectrum changes significantly, especially during the second
event when the temperature of the plasma was in excess of 5 keV. 
Evidence of such high temperatures is provided by the appearance of the complex { of} lines 
around 6.7 keV owing to highly ionized Fe. 

\begin{figure}
\resizebox{\columnwidth}{!}{
                \includegraphics{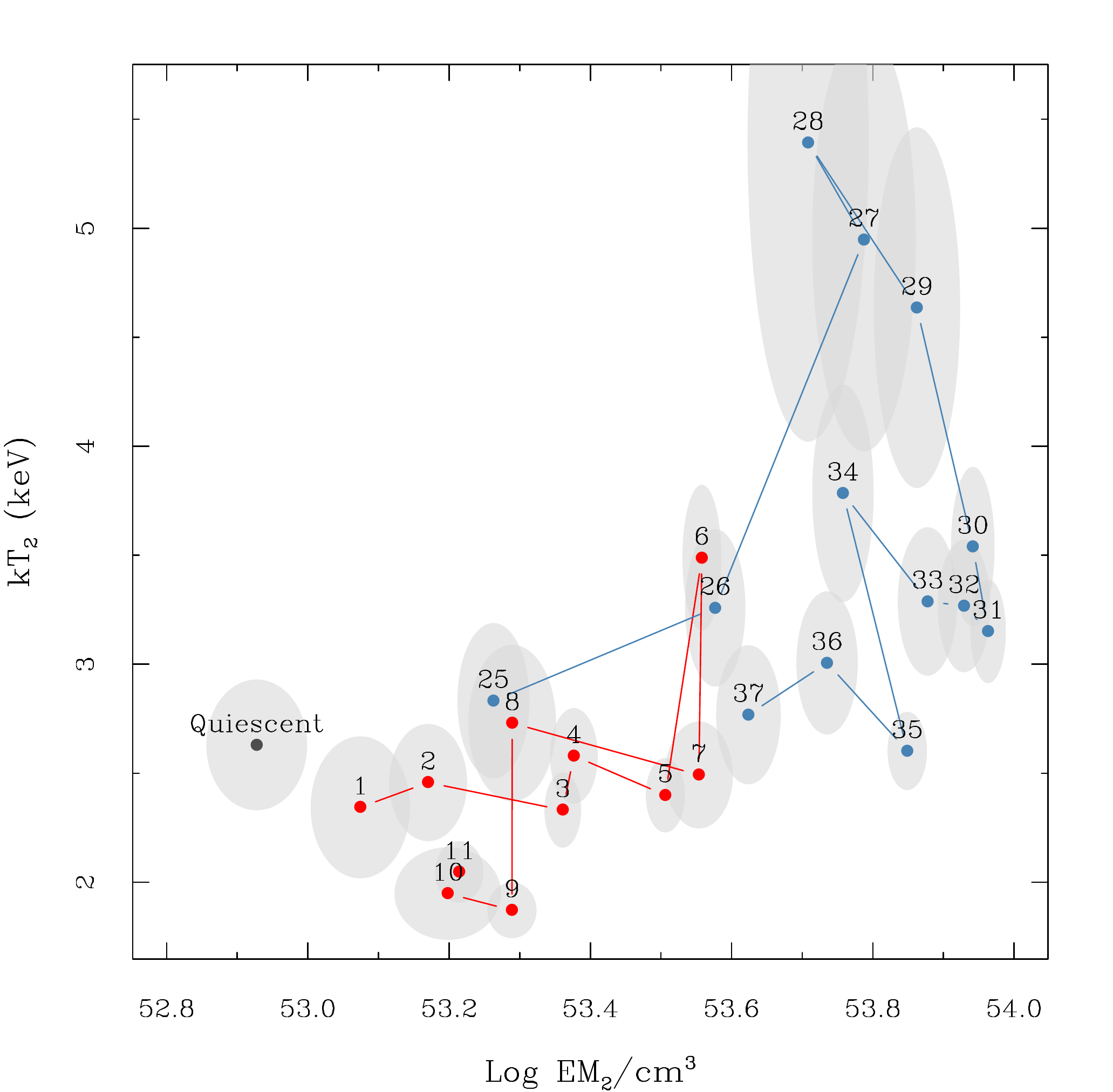}
                }
\caption{\label{logtem} $kT_2$ vs. emission measure ($EM_2$) of the hot component of the 2T APEC 
best-fit models for the time intervals (see text) relative to the first event (red points)\ 
and second event (blue points).
The point relative to the quiescent phase (time intervals $12-24$) corresponds to the hottest
component of the 3T VAPEC best-fit model. Ellipses denote the $1\sigma$ confidence ranges.}
\end{figure}

\noindent{{\em First event} (intervals $1-10$).} 
The hot temperature during this event varied by 1.9 keV$<kT<$2.5 keV (Table \ref{tabfit},
Fig. \ref{logtem}) apart from interval number 6 when it had a sudden increase ($kT \sim3.4$ keV) 
of about 3$\sigma$ higher than the average of the previous values such heating rapidly vanished 
during the next interval (number 7). 
{The spectrum corresponding to the intervals around the peak of the rate increase is shown in Fig. \ref{epicspectra} 
(red curve).}
During intervals 8 through 10 the hot temperature 
slightly cooled down toward values lower than those seen at the beginning of the observation.
We can conclude that, on average, only the plasma emission measure (EM) increased significantly 
during intervals $2-7$, while the temperature  was almost steady. 
This behavior fits the scenario in which a hot spot gradually appears on the stellar surface 
due to the stellar rotation  (P$_\mathrm{rot}\sim1.205$ days $\sim$ 104.1 ks).
In interval number 6 either a short flare happened or the very hot core of the region 
was visible for a short time.
The vanishing of the high rate is plausible with the gradual disappearance of the same spot 
at the opposite limb of the star. 

\noindent{\em Quiescent state.} 
During intervals $11-24$ the star showed a relatively small flare during interval
number 13 ($\sim55$ ks since the start of the observation, visible in four bins of 
the light curve of Fig. \ref{lcpn}) and with a duration of approximately 5 ks, 
plus other small scale variability afterwards. This demonstrates that even during the 
{\em quiescent} state some degree of X-ray variability was present.
When modeling the plasma with two absorbed thermal APEC components, 
the cool component was around $0.8-0.9$ keV and the hot component in { 1.9 keV $< kT <2.7$ keV}, 
which is within 2$\sigma$ of the values of hot temperatures seen during the first event. 
On average, the quiescent flux is about $1.35\times10^{-12}$ \fxu, which is very similar
to the quiescent flux measured in 2013 ($1.5\times10^{-12}$ \fxu).

 We grouped together the time bins between 11 and 24 (interval 13 excluded) 
to gain more count statistics and allow a more refined analysis of the corresponding spectrum
(black curve in Fig. \ref{epicspectra}). 
We tried models with two, three, and four APEC components and models with two and three VAPEC components. 
The 2T APEC model does not give us a statistically valid fit, 
while the three APEC component model does. The model with four APEC components does not improve the 
fit to data.  For the three APEC components we find 
$kT_1=0.26\pm0.03$ keV, $kT_2=0.98\pm0.04$ keV, and $kT_3=2.97\pm0.7$ keV with ratios of 
emission measures (EM) $EM_3/ EM_1 \sim 0.12$, $EM_2/EM_1 \sim 0.4$.
We thus detect a soft component at 0.26 keV and there is evidence of a hot component at $\sim3$ keV
also during quiescence. 

{ For the VAPEC models we left Fe, O, and Ne free to vary and linked all other elements to the Fe abundance.}
Both the 2T VAPEC model and the 3T VAPEC give a satisfactory fit to the data.
For the 2T VAPEC model the two temperatures are  $kT_1 = (0.81\pm0.04)$ keV and $kT_2 =(2.6\pm0.5)$ keV,
respectively, with $EM_2/EM_1 \sim 0.5$. The abundances of Fe and other metals are found around 
$0.13Z_\odot$, but with O and Ne around 1.1 and 0.75 times the solar value. This pattern is 
similar to that observed in active stellar coronae (the so-called inverse FIP effect, 
see, e.g., \citealp{Guedel2004} and references therein) where the abundance of Fe (an element with low 
first ionization potential, FIP) is found depleted with respect to the value measured in the 
solar corona, while the abundances of elements with high FIP, such as O and Ne, appear enhanced.

For the 3T VAPEC model, a pattern of abundances similar to that of the 2T VAPEC model fit
is found ($Fe/Fe_\odot = 0.15$, 
$[O/Fe] \sim 1.3$, $[Ne/Fe]\sim0.8$). In this case the temperatures are $kT_1 = (0.1\pm0.1)$ keV,
$kT_2 = (0.80\pm 0.04)$ keV, and $kT_3 = (2.63\pm0.6)$ keV with EM ratios of $EM_2/ EM_1 \sim 4.5$, 
$EM_3/EM_1 \sim 1.9$. We detect thus a cool component and a hot component in the quiescent 
phase spectrum, whereas the main component remains that at $\sim0.8$ keV.
This also justifies the use of a simple 2T APEC model for time resolved spectroscopy,
and the focus on the hot component of such a model to study the two main increases of X-ray 
flux of \rhoOph~A.

\noindent{{\em Second event} (flare, intervals $25-37$).}
These intervals contain the second and most powerful event observed in \rhoOph~A.
The \pn\ spectrum relative to the time bins around the peak of the rate is shown in
Fig. \ref{epicspectra} (green curve). The spectrum is characterized by the appearance of lines
of highly ionized Fe at 6.7 keV.
The rise of the flare started at interval number 25, 
and its evolution is best described by the hot component of the spectral fitting, which we discuss here.
An increase of EM and temperature is detected, starting from intervals 25 and 26,
and even more during intervals $27-29$. Interval number 28 shows the peak of $kT\sim5.4$ keV,
then  $kT$ decreases steadily through intervals $29-31$, and more slowly through intervals $32-37$, 
with some apparent reheating at interval 34. 
At these times EM showed its maximum values (with a peak of $\log EM (\mathrm{cm^-3})\ge 53.9$ in 
interval 31) then it decreased toward pre-flare values. The peak of the hot temperature was reached in interval 28 
and does not coincide with the maximum of EM, which peaked during interval  31. 
This fact is consistent with what is observed in the flares of cool stars and in the Sun, where the 
initial heating and peak of temperature is followed by an increase of the density of the flaring loop, 
filled up by plasma evaporating from the loop footpoints \citep{Reale2007}.

The behavior of the flare closely resembles that of solar and stellar flares observed in X-rays,
where energy is suddenly released with an impulse in plasma magnetically confined in loops,
followed by a longer cooling decay. We presume that \xmm\ did not completely observe the full decay of 
the flare and the disappearance of the hot region as in the first event. 
This and the too shallow track of the decay in the $\rm{EM}-\rm{kT}$ 
diagram did not allow us to have reliable 
diagnostics from the flare decay as in \citet{Reale1997}, and we preferred to derive loop diagnostics from 
the analysis of the rise phase, as described in \citet{Reale2007}. The loop length can be derived  
from measuring  the duration of the rise of the light curve and/or the time delay of the emission measure peak 
from the temperature peak \citep{Reale2007} as follows:
\begin{equation}
L_9 \approx 3 ~ \psi^2 T_{0,7}^{1/2} t_{M,3} \approx 190
\label{eq:lris}
\end{equation}
\begin{equation}
L_9 \approx 2.5 ~ \frac{\psi^2}{\ln \psi} T_{0,7}^{1/2} \Delta t_{0-M,3} \approx 140
\label{eq:ldelt}
,\end{equation}
where
\[
\psi = \frac{T_0}{T_{M}} \approx 1.9,
\]
$L_9$ is the loop half-length in units of $10^9$ cm, $t_{M,3} \approx 4.6$ ks is the time of the emission 
measure maximum measured from the flare beginning, $T_{0}$ ($T_{0,7}$) the loop maximum temperature (in units of $10^7$ K),  
$T_{M}$ the temperature at the emission measure maximum, and $\Delta t_{0-M,3} \approx 2.6$ the time elapsed between the 
peak of the temperature and the peak of the emission measure in ks. The values $T_0 \approx 150$ MK (bin 28) and $T_M \approx 80$ MK (bin 31) 
are maximum loop temperatures inferred from the measured temperatures, which average over the whole flaring loop
(see Appendix in \citealt{Reale2007}). In summary,
we inferred a loop half-length $L= 1.4-1.9\times10^{11}$ cm, which corresponds to about $25\%-33\%$ of the stellar radius
or about $2-2.7$ times the solar radius. 
We also estimated that the coronal magnetic field must have an intensity $B \ge 300$ G to confine the plasma 
within the loop.

\section{RGS spectra}
\begin{figure*}
\resizebox{\textwidth}{!}{
   \includegraphics[width=0.48\textwidth,angle=0]{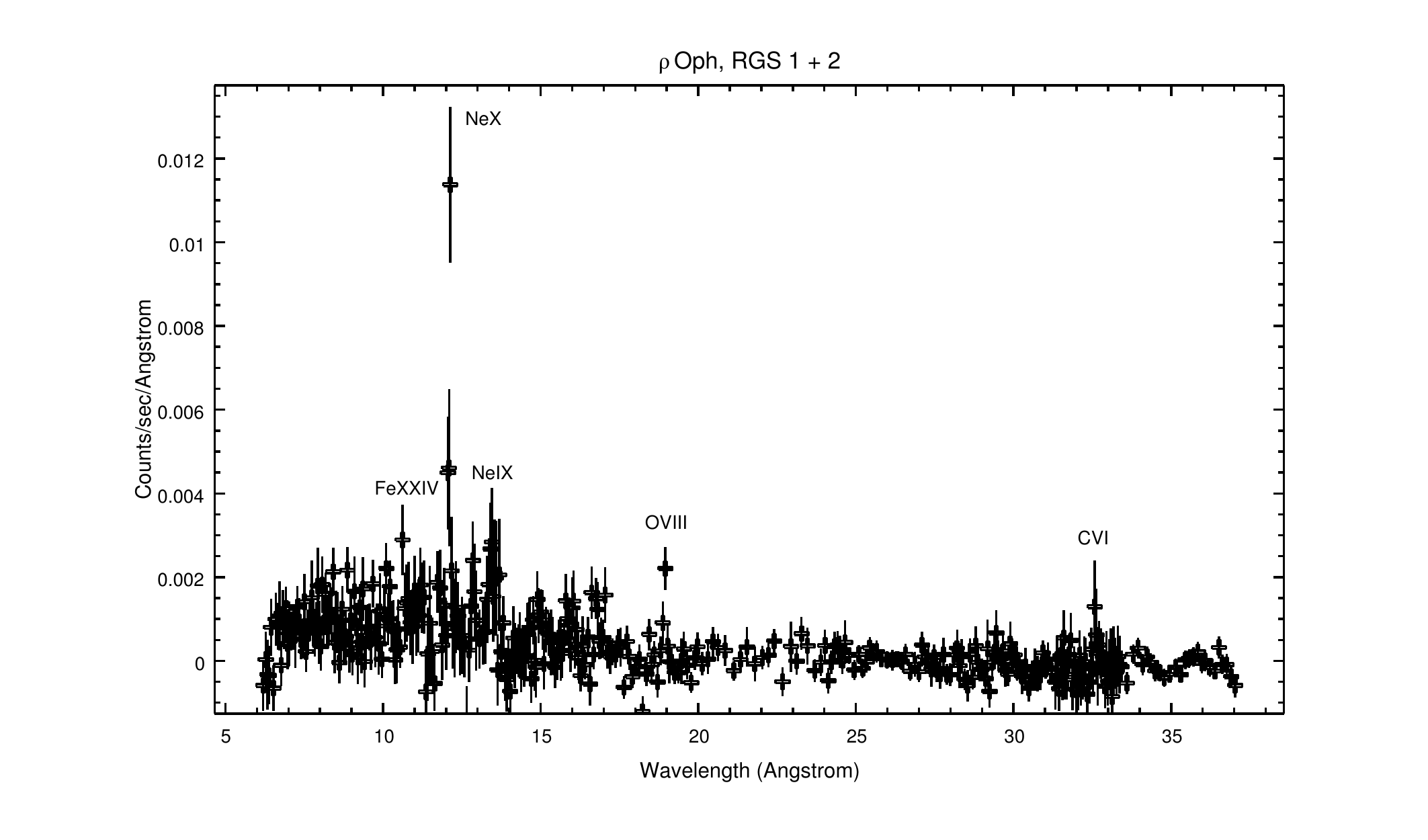}
   \includegraphics[width=0.48\textwidth]{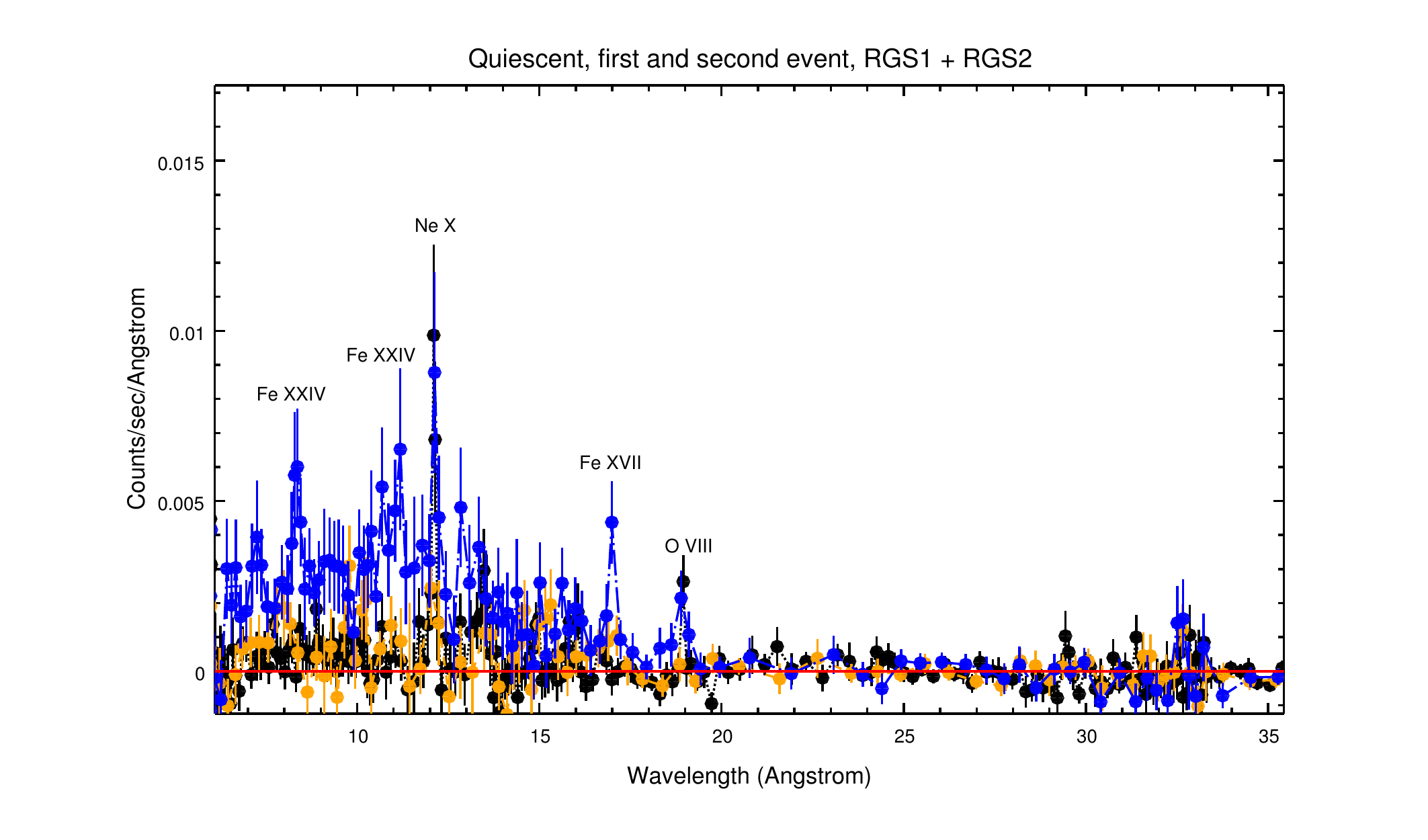}
}
\caption{\label{rgs} RGS spectra of \rhoOph~A in the range $5-30$\AA. 
Left panel: the average spectrum during the full exposure (140 ks). The spectrum 
was binned to have at least 25 counts per bin. Right panel: 
black refers the spectrum relative to the quiescent state between the two main variability events, 
the orange spectrum refers to the first event, and the blue spectrum refers to
the second event. Changes in the intensities of lines are visible  mostly during the flare 
event at wavelengths below 12\AA, when both lines and continuum increased their strength.}
\end{figure*}

The average spectrum of \rhoOph~A is shown in  Fig. \ref{rgs} (left panel), while
the spectra of the quiescent ($40-115$ ks, black) interval,  the first ($10-40$ ks, orange) 
and second ($115-140$ ks, blue) 
event are shown in the right panel of Fig. \ref{rgs}.
These spectra are rebinned to have at least 25 counts per bin to enhance the signal, 
although this reduces the original spectral resolution. 
Qualitatively, hot lines from \ion{O}{VIII} and \ion{Ne}{X} are detected both in the quiescent interval 
and during the variability events, indicating the presence of hot plasma ($kT \ge 1$ keV) that is already
detected in the \pn\ spectra. 
Another sign of high temperature plasma is the relatively high level of continuum,  
below 10\AA, similar to what has been observed in stars with high activity and plasma
temperatures like, for example, UX Ari and $\epsilon$ Eri \citep{Ness2002}. 

The best fit of the average RGS spectrum with two absorbed  APEC thermal components results in 
a temperature around 0.3 keV and a hot temperature around 5 keV for a fixed $Z=0.3Z_\odot$, in agreement
with the \pn\ spectral analysis, even though the range of energy of the RGS is limited to $\sim 2.5$ keV.

Changes of line intensities are mostly observed during the second event. In particular, 
we observed an increase of both continuum and line strength at wavelengths shorter than 12\AA\ during
the second (flare) event. Lines of \ion{Fe}{XXIV} are visible during the flare owing to temperatures in excess of
5 keV as determined from the analysis of the \pn\ spectra. 

We measured the line fluxes by modeling single lines with Gaussian profiles characterized by a line central
wavelength and a full width at half maximum (FWHM). In this way we take into account any intrinsic widths 
of the lines. The model also comprises a continuum level in a window of about $0.7-1$\,\AA\ 
around each feature; we used CIAO Sherpa 4.8\footnote{\url{http://cxc.harvard.edu/sherpa4.8/} 
to obtain the best fit parameters.}  
Table \ref{lines} presents the measurements of the main lines identified in the spectra. 
Determining a 1$\sigma$ confidence range was difficult for some lines as it depends on the statistics 
and number of bins available for fitting. As a consequence, any quantitative
conclusions on the widths of the lines and the amount of change during the increase of rates are 
impossible with these data. In general the line widths are consistent with zero even for the
spectra relative to the high rate events, when we presume that a line broadening should be present owing to the traveling of the source of X-rays across the surface, either this is an active spot or an unseen 
companion during the interval in which the spectra  were accumulated. 
Qualitatively, RGS spectra show the potential for a deeper high resolution spectroscopic 
follow up of \rhoOph\ in order to improve the statistics and precisely measure line widths and 
velocity fields.  

\section{Discussion and conclusions}\label{discussion}
We have presented an \xmm\ observation that monitored \rhoOph\ for 140 ks, 
encompassing  an entire rotation period of the primary star of the system ($P\sim 1.2$ days). 
We observed a change of X-ray emission twice, each time lasting about $35$ks and separated by 
about 104 ks.
This ``lighthouse'' effect has increased its strength in the 2016 observation. On the other hand,
the quiescent phase has remained at similar levels of emission with respect to the 2013 observation.
The cause of this phenomenon can be attributed either to a large active spot 
on the stellar surface, or an unknown low mass companion, and both scenarios have strengths
and weaknesses.

\noindent{\bf Active spot.} In this scenario, the spot is a magnetospheric feature that could also have
a photospheric counterpart and has been steadily present on the stellar 
surface for at least 2.5 yrs (i.e., since the epoch of the 2013 \xmm\ observation, Paper I),
as the light curves of 2013 and 2016 can be put in phase with the stellar rotation period,  
thus matching the  periodic increases of rate. In this respect \rhoOph~A would be a magnetic
early B star similar to those discovered by, for example, \citet{Fossati2014}.
From the duration of the rate change (about $30-35$ ks) compared to the period (104 ks) 
and taking into account the inclination angle of the star ($i\sim45\degr$),
we infer a quite large size for the spot, amounting to about 1.5 stellar radii. 
The cusp-like shape of the first episode of variability, free of major flares, 
and the lack of a flat top level suggest that such a large spot could be plausible 
because when the spot is completely on view after emerging from one limb it starts to disappear
at the opposite limb.

We attribute the origin of the spot to a strong magnetic field that is responsible for 
creating the spot in the form of magnetic loops. The loops confine high temperature plasma that emits
X-rays. In this scenario the spot can be seen as a scaled-up version of loop-populated solar active regions. 
Flares can occur within these loops that make the typical temperature of the plasma 
rise above $kT\ge5$ keV, as observed in the second event. 
No constraints on the magnetic field
in \rhoOph~A could be found in the literature, therefore 
magnetic field scenario remains hypothetical at the moment. In the future work we aim to obtain reliable magnetic field measurements of this interesting object.

If \rhoOph~A has a magnetic field, then it would not be the only known early B star exhibiting surface 
spots and a magnetic field.
In $\tau$~Sco \citet{Donati2006} discovered a magnetic field with a complex topology.
The magnetic loops should be responsible for emitting X-rays as in solar-type stars.
However, the lack of modulated X-ray emission hints that the magnetic loops are small and 
distributed across the surface \citep{Oskinova2016,Ignace2010}.  
However, in \rhoOph\ we find well-defined periodic modulation, thus a strong magnetic field 
must be concentrated in a large spot. We speculate that the magnetic field has a large scale
structure that is similar to a simple dipolar configuration and a stable configuration deduced by 
the long lifetime of the spot  ($\ge 2.5$ yrs). 

\citet{Wade2016} have found a strong magnetic field, on the order of a few kG, in HD 164492C1, which is 
an early B star similar to \rhoOph\ rotating in 1.36 days and part of a hierarchical system with other
two A/Herbig Ae stars belonging to the Trifid nebula. However, one difference from \rhoOph\  is that the 
HD 164492 system is tighter than that of \rhoOph\ and the origin of the magnetic field in such objects 
could be associated with binarity and the stellar formation mechanism.
Another case of strong magnetism in an early B star (CPD~$-62\degr 2124$) is reported by \citet{Hubrig2017},
where a strong dipolar magnetic field with strength in excess of $\sim30$ kG has been detected.
Analogous to \rhoOph~A, CPD~$-62\degr 2124$ is a fast rotator with rotational period of about 1.45 days.

Weak emission of X-rays has been detected in $\sigma$~Sgr (B2V, $\sim65$ pc) in a short 10 ks 
\xmm\ observation (PI Oskinova; XMM ObsId 0721210101). 
For this star we derived a flux of about $7.4\times10^{-15}$ \fxu\ in 0.3-8.0 keV and a luminosity
of $\sim3.7\times10^{27}$ \lxu, which is about three orders of magnitude lower than the X-ray 
luminosity of \rhoOph~A. The count statistics are low but we can infer that the spectrum is almost
entirely below 1 keV and peaks around 0.7 keV, pointing to a substantial difference in the
mechanisms that  produce X-rays in $\sigma$~Sgr and \rhoOph. 

B stars do not possess magnetic coronae { such as solar-type stars}, 
and in this respect they are similar to A type stars,
which are mostly dark in X-rays and lack strong magnetic fields. However, even among 
A type stars, which are presumed to be less active stars during the main sequence lifetime  
because they lack a solar-type corona, { \citet{Balona2013} found a number of stars with spots 
in an analysis based on {\em Kepler} data.}

\noindent{\bf Low mass companion.} 
Another explanation is the presence of an unknown low mass
stellar companion { as the source of X-rays}. 
If such a companion is fully responsible for the X-rays, it means that it is not completely eclipsed 
by \rhoOph~A when orbiting around it. To support this we notice that \rhoOph~B (another B2 star) 
is undetected in the same data. During the quiescent state we observed the less active part 
of the corona of the unknown companion, while during the recurrent increase of rate 
we had observed the most active part of it, 
i.e., perhaps the part of the surface of the companion facing the primary (its day side).
From the combined analysis of the X-ray derived period and the v~sin$i$ from the line profile,
it is suggested that the primary is inclined by $\sim45\degr$. If the plane of the orbit of the
companion has the same inclination it is plausible that during the superior conjunction part of its
surface (its day side) is partially obscured by the primary, dark in X-rays.  

Two factors can boost the X-ray activity of the hypothetical companion to such a high level: 
its young age, coeval with \rhoOph\ and with the
other solar-type stars discovered around it \citep{Pillitteri2016}, and being part of a tight
binary system. This would enhance the stellar activity even more than in a configuration of single star, 
as observed for example in the RS CVn systems. 
In fact, if the companion rotates in 1.2 days, as \rhoOph s does, such a period would correspond  
to a very close separation of $\sim9.8 \mathrm{R}\odot$ that is similar to the radius of \rhoOph~A. 
This results in an extreme system with the companion almost on the verge of collapsing 
onto the primary. The system should be in a 1:1 spin-orbit locked configuration reached 
in the first 5 Myr, making its evolution fast. 

The plasma temperatures, the type of impulsive variability and the pattern of abundances 
of the quiescent spectrum are fully consistent with the X-ray properties of 
young, active pre-main sequence stars.
If a young unseen companion emits such X-rays, the derived loop length ($1.4-1.9\times10^{11}$ cm 
or $\sim2-2.7 R_\odot$) would be a few of its stellar radius. 
This implies an extended structure that perhaps could connect the surface of \rhoOph~A 
with the companion and hints to some form of magnetic interaction between the two stars. 
At the same time the hottest part of the corona of the companion should be eclipsed at some 
phases of the orbital motion to reproduce the features observed in the light curve.

In the literature we find cases of X-rays from massive stars that are attributed to a low mass companion,
such as $\sigma$ Ori E (B2Vp; \citealp{Sanz-Forcada2004}), which was detected in a flaring state during a \xmm\ observation.
$\sigma$ Ori E has a hard spectrum similar to \rhoOph~A and remarkably harder than that of $\sigma$ Ori AB 
(O9.5V). However, in contrast to $\sigma$ Ori E we did not detect a 6.4 keV line from neutral Fe, 
even during interval { 28} when the spectrum reached the maximum temperature. 
The absence of a circumstellar disk on the companion of \rhoOph~A is perhaps linked to the lack of 
the 6.4 keV line as seen in $\sigma$ Ori E. There, the neutral Fe could emit by fluorescence 
during the flare as the relatively cold material of the disk is hit by high energy X-ray photons 
created during the flare, while in \rhoOph~A  there is no circumstellar disk from where 
fluorescence of neutral Fe can arise.

The analysis of the optical spectra from the ESO archive did not reveal any Doppler shift 
due to the presence of the companion. This does not exclude its presence but rather calls for 
an ad hoc spectroscopic campaign on \rhoOph~A. 

A mixture of both scenarios could be present as well; that is, X-rays produced by the coronal 
activity of an unseen companion coupled to and enhanced by an intrinsic magnetism of
\rhoOph~A. 

Among other cases of uncertain origin of X-rays we cite IQ Aur, which is an A0p 
bright in X-rays and flaring during a \xmm\ observation with temperatures up to 6 keV
\citep{Robrade2011}. In the case of IQ Aur a magnetized wind model and the presence of an unseen
low mass companion are discussed by Robrade et al. Qualitatively the X-rays from  IQ Aur are 
similar to those from \rhoOph\ except for the fact that the flare has been observed once and 
never with periodic recurrence. In the case of a magnetized wind emission the occurrence of 
a flare does not easily fit the current models. 

Other mechanisms of production of X-rays observed in binary systems of 
massive stars are less probable because the separation between \rhoOph~A and B is about 300 AU,  the short period
of the phenomenon appears to be linked to the rotation of the primary, and because the winds from
B2 stars are weaker than those from O and Wolf Rayet stars. 

A clear case of periodic variation of X-ray flux of a massive star has been reported 
in $\chi^1$ CMa by \citet{Oskinova2014}. 
$\chi^1$ CMa is a variable Beta Cep type star (B0.5-1 V-IV) characterized
by a strong magnetic field ($B>5000$ G). $\chi^1$ CMa  exhibits X-ray and H-band mag variability 
in phase, however its X-ray spectrum is softer than \rhoOph\ and it does not show flares. 
The mechanism that produces X-rays in $\chi^1$ CMa is still not completely clear; 
it is likely linked to the compression phase and the plasma heating in the consequent shock, as the 
maximum rate of X-rays happens at the minimum stellar radius. This scenario is not applicable
to \rhoOph\ because it cannot explain the much higher plasma temperatures observed in it and its flaring 
activity. 
 
The present work indicates the peculiarity of \rhoOph\ in the X-ray band,
as it represents perhaps the best example of an X-ray active early B star and 
one of the most favorable targets for studying magnetism in B stars. 
On the other hand, \rhoOph~A could be an extreme system with a low mass companion on 
the verge of collapsing onto the primary. In both cases \rhoOph~A deserves more investigation
to understand the origin of its X-rays.

\begin{acknowledgements}
We would like to thank the anonymous referee for constructive comments.  
IP is grateful to dr. Javier Lopez-Santiago and dr. Ines Crespo-Chacon for providing useful information 
on the analysis of the line profile and the derivation of v~sin$i$.
S.J.W. was supported by NASA contract NAS8-03060.
\end{acknowledgements}

\begin{appendix}
\section{Time resolved spectroscopy}
\begin{table*}[t]
\centering
\caption{
Best-fit parameters from time resolved spectroscopy of the first event and second event.
We list the temporal intervals, temperatures, 
logarithm of the emission measures (EM), unabsorbed fluxes in bands  $0.3-1.0$ keV ($F_s$), 
$1.5-8.0$ keV ($F_h$), $0.3-8.0$ keV ($F_{tot}$), unabsorbed X-ray luminosity in $0.3-8.0$ keV, hardness ratio ($HR = F_h/F_s$), and  $\chi^2$ statistics. 
}
\label{tabfit}
\begin{center}
\resizebox{1.05\textwidth}{!}{
\begin{tabular}{rrr|rrrrrr|rrrrr|rrr}
  \hline
  \hline
Interval & Start & End & T1 & Err (T1) & T2 & Err (T2) & EM1  Err (EM1) & EM2  Err (EM2) & $F_s$ & $F_h$ & $F_{tot}$ & $\log L_X$ & HR & $\chi^2$ & d.o.f. & $P(\chi^2>\chi^2_0)$ \\ 
First Event& ks & ks & \multicolumn{4}{c}{keV} &  \multicolumn{2}{c|}{$\log$ cm$^{-3}$} & \multicolumn{3}{c|}{$10^{-13}$\fxu} & \lxu & & & & \\
  \hline
  1 & 0.00 & 5.45 & 0.89 & 0.06 & 2.34 & 0.33 & 52.94  $_{-0.07} ^{0.06}$    & 53.00  $_{-0.08} ^{0.06}$ & 7.66 & 4.77 & 15.63 & 30.36 & 0.62 & 1.22 &  36 & 16.75 \\ 
    2 & 5.45 & 10.76 & 0.87 & 0.06 & 2.46 & 0.27 & 52.90  $_{-0.07} ^{0.06}$ & 53.09  $_{-0.06} ^{0.05}$ & 8.00 & 5.80 & 17.19 & 30.40 & 0.73 & 1.25 &  38 & 13.96 \\ 
    3 & 10.76 & 15.20 & 0.61 & 0.05 & 2.33 & 0.17 & 52.90  $_{-0.05} ^{0.04}$ & 53.29  $_{-0.03} ^{0.03}$ & 10.17 & 7.73 & 21.92 & 30.51 & 0.76 & 1.12 &  38 & 28.36 \\ 
    4 & 15.20 & 19.04 & 0.83 & 0.05 & 2.58 & 0.22 & 52.95  $_{-0.05} ^{0.04}$ & 53.30  $_{-0.04} ^{0.03}$ & 10.67 & 9.22 & 24.49 & 30.56 & 0.86 & 1.35 &  38 & 7.19 \\ 
    5 & 19.04 & 22.44 & 0.73 & 0.07 & 2.40 & 0.17 & 52.85  $_{-0.07} ^{0.06}$ & 53.43  $_{-0.03} ^{0.03}$ & 11.64 & 11.05 & 27.93 & 30.61 & 0.95 & 1.61 &  38 & 0.98 \\ 
    6 & 22.44 & 25.05 & 0.81 & 0.06 & 3.49 & 0.33 & 53.03  $_{-0.05} ^{0.05}$ & 53.48  $_{-0.03} ^{0.03}$ & 14.27 & 17.20 & 37.65 & 30.74 & 1.21 & 1.17 &  40 & 21.82 \\ 
    7 & 25.05 & 27.74 & 0.89 & 0.08 & 2.49 & 0.25 & 53.06  $_{-0.1} ^{0.08}$ & 53.48  $_{-0.05} ^{0.05}$ & 14.61 & 13.43 & 34.81 & 30.71 & 0.92 & 1.13 &  39 & 26.03 \\ 
    8 & 27.74 & 31.24 & 0.95 & 0.04 & 2.73 & 0.36 & 53.16  $_{-0.06} ^{0.05}$ & 53.21  $_{-0.07} ^{0.06}$ & 11.79 & 8.99 & 26.14 & 30.59 & 0.76 & 1.89 &  36 & 0.10 \\ 
    9 & 3:1.24 & 36.05 & 0.79 & 0.05 & 1.87 & 0.13 & 52.90  $_{-0.05} ^{0.05}$ & 53.21  $_{-0.04} ^{0.03}$ & 9.36 & 5.70 & 19.20 & 30.45 & 0.61 & 0.88 &  36 & 66.73 \\ 
   10 & 36.05 & 41.22 & 0.88 & 0.09 & 1.95 & 0.21 & 52.91  $_{-0.1} ^{0.09}$ & 53.12  $_{-0.08} ^{0.07}$ & 8.33 & 5.09 & 17.18 & 30.40 & 0.61 & 1.56 &  35 & 1.91 \\ 
   11 & 41.22 & 46.68 & 0.74 & 0.05 & 2.05 & 0.14 & 52.86  $_{-0.05} ^{0.04}$ & 53.14  $_{-0.04} ^{0.03}$ & 8.34 & 5.18 & 16.91 & 30.40 & 0.62 & 1.48 &  35 & 3.28 \\  \hline
Second Event&       &       &      &      &      &      &       &       &       &       &      &      &       &       &      &            \\ 
   25 & 115.76 & 120.09 & 0.90 & 0.06 & 2.83 & 0.36 & 52.98  $_{-0.07} ^{0.06}$ & 53.19  $_{-0.05} ^{0.05}$ & 9.47 & 8.05 & 21.65 & 30.50 & 0.85 & 1.14 &  38 & 26.03 \\ 
   26 & 120.09 & 122.48 & 1.01 & 0.05 & 3.26 & 0.36 & 53.13  $_{-0.08} ^{0.07}$ & 53.50  $_{-0.04} ^{0.04}$ & 14.54 & 17.91 & 39.91 & 30.77 & 1.23 & 1.15 &  40 & 24.14 \\ 
   27 & 122.48 & 123.99 & 1.26 & 0.13 & 4.95 & 0.97 & 53.43  $_{-0.2} ^{0.1}$ & 53.71  $_{-0.08} ^{0.07}$ & 20.61 & 37.10 & 69.06 & 31.01 & 1.80 & 0.81 &  43 & 81.24 \\ 
   28 & 123.99 & 125.29 & 1.22 & 0.08 & 5.39 & 1.37 & 53.59  $_{-0.1} ^{0.09}$ & 53.63  $_{-0.09} ^{0.08}$ & 24.51 & 36.62 & 75.48 & 31.05 & 1.49 & 1.21 &  39 & 17.76 \\ 
   29 & 125.29 & 126.57 & 1.11 & 0.18 & 4.64 & 0.83 & 53.35  $_{-0.2} ^{0.1}$ & 53.79  $_{-0.07} ^{0.06}$ & 24.07 & 42.29 & 79.74 & 31.07 & 1.76 & 0.96 &  43 & 54.67 \\ 
   30 & 126.57 & 127.81 & 0.91 & 0.09 & 3.54 & 0.37 & 53.20  $_{-0.1} ^{0.09}$ & 53.87  $_{-0.03} ^{0.03}$ & 27.33 & 41.33 & 82.12 & 31.08 & 1.51 & 1.05 &  41 & 37.73 \\ 
   31 & 127.81 & 129.03 & 0.78 & 0.08 & 3.15 & 0.24 & 53.20  $_{-0.08} ^{0.06}$ & 53.89  $_{-0.03} ^{0.02}$ & 29.77 & 39.43 & 83.03 & 31.09 & 1.32 & 0.83 &  40 & 76.04 \\ 
   32 & 129.03 & 130.35 & 1.02 & 0.06 & 3.27 & 0.30 & 53.34  $_{-0.1} ^{0.08}$ & 53.85  $_{-0.04} ^{0.04}$ & 28.62 & 39.28 & 82.93 & 31.09 & 1.37 & 0.91 &  42 & 62.93 \\ 
   33 & 130.35 & 131.57 & 1.03 & 0.05 & 3.29 & 0.34 & 53.46  $_{-0.08} ^{0.06}$ & 53.80  $_{-0.04} ^{0.04}$ & 29.63 & 36.34 & 81.30 & 31.08 & 1.23 & 0.84 &  39 & 75.48 \\ 
   34 & 131.57 & 132.93 & 0.93 & 0.05 & 3.78 & 0.50 & 53.49  $_{-0.06} ^{0.05}$ & 53.68  $_{-0.05} ^{0.04}$ & 29.04 & 30.65 & 72.59 & 31.03 & 1.06 & 0.82 &  40 & 78.41 \\ 
   35 & 132.93 & 134.49 & 0.79 & 0.07 & 2.60 & 0.18 & 53.18  $_{-0.07} ^{0.06}$ & 53.77  $_{-0.03} ^{0.03}$ & 25.03 & 26.31 & 62.84 & 30.97 & 1.05 & 1.00 &  41 & 47.05 \\ 
   36 & 134.49 & 136.14 & 1.04 & 0.05 & 3.01 & 0.33 & 53.33  $_{-0.08} ^{0.06}$ & 53.66  $_{-0.05} ^{0.04}$ & 21.66 & 24.75 & 57.82 & 30.93 & 1.14 & 1.10 &  39 & 30.78 \\ 
   37 & 136.14 & 138.19 & 0.98 & 0.05 & 2.77 & 0.32 & 53.29  $_{-0.07} ^{0.06}$ & 53.55  $_{-0.05} ^{0.04}$ & 18.98 & 18.28 & 46.58 & 30.84 & 0.96 & 1.19 &  38 & 20.05 \\ 
   \hline
\end{tabular}
}
\end{center}
\end{table*}

\begin{table*}
\caption{\label{lines} Lines measured in the RGS spectra relative to the full exposure, 
quiescent, and high state. For each line, Gaussian FWHM (\AA), line positions (\AA) and 
intensities (ct s$^{-1}$\AA$^{-1}$) are indicated.}
\begin{center}
\resizebox{0.47\textwidth}{!}{
\begin{tabular}{lrccc} \hline
\hline
Ion &   &       Full & Quiescent & High State \\ 
\hline
\ion{Fe}{XXIV} &  FWHM (\AA)  & $0.004_{-0.004} ^{0.288}$   & -- & $0.08_{-0.08} ^{0.17}$ \\
8.0\AA\        &  pos. (\AA)  & $8.08_{-0.04} ^{0.40}$         & -- & $8.35_{-0.12} ^{0.04}$  \\
               &  int. (ct/ks) &  $0.6_{-0.6} ^{0.6}$  & -- & $0.11_{-0.08} ^{3.67}$ \\
\hline
\ion{Fe}{XXIV} & FWHM  (\AA) &  $0.011_{-} ^{0.061} $   & $0.01{-} ^{0.22}$      &   --  \\ 
10.6\AA\ & pos.  (\AA) &  $10.64_{-0.017} ^{0.016}$   & 10.64  --        &   --  \\
         & int. (ct/ks) &  $0.33_{-0.3} ^{1.4}$   & $0.04_{-0.15} ^{0.15}$ &   --  \\
\hline
\ion{Ne}{X} & FWHM  (\AA)  &  $0.0016_{-0.0004} ^{0.0442}$ & $0.012_{-} ^{0.05}$ &  $0.005_{-0.003} ^{0.005}$ \\
12.1\AA\     & pos.  (\AA) &  $12.12_{-0.01} ^{0.01}$    &  $12.11_{-0.01} ^{0.01}$     &  $12.13_{-0.01} ^{0.01}$ \\
             & int.(ct/ks) &  $6.0_{-5.0} ^{8.0}$    &  $0.82_{-0.07} ^{1.63}$ &  $1.5_{-1.3} ^{-}$  \\
\hline \\
\ion{Ne}{IX} & FWHM (\AA)&  $0.10_{-0.1} ^{0.04}$   & $0.29_{-0.28} ^{0.24}$  &  $0.004_{-} ^{0.14}$ \\
13.4\AA\   & pos. (\AA)  &  $13.44_{-0.05} ^{0.05}$ & $13.43_{-0.09} ^{0.09}$   &  $13.42_{-} ^{-}$  \\
          & int.(ct/ks)   &  $0.036_{-} ^{3.0}$   & $0.026_{-0.014} ^{10.000}$   &  $0.39_{-1.4} ^{3.5}$ \\
\hline \\
\ion{Fe}{XVII} & FWHM (\AA) &  $0.10_{-0.10} ^{0.04}$        &  $0.15_{-0.15} ^{0.10}$        &  $0.001_{-} ^{0.15}$ \\
15\AA\       & pos. (\AA) & $14.97_{-0.07} ^{0.05}$       &  $14.92_{-0.07} ^{0.06}$       & $15.03_{-0.03} ^{0.04}$ \\
             & int. (ct/ks) &  $0.04_{-0.03} ^{4.6}$ &  $0.03_{-0.02} ^{27.00}$ & $4.0_{-1.1} ^{3.0}$ \\
\hline \\
\ion{O}{VIII} & FWHM  (\AA)  &  $0.07_{-0.07} ^{0.03}$  & $0.06_{-0.06} ^{0.02}$ &  $0.14_{-0.14} ^{0.07}$ \\
18.9\AA    &  pos. (\AA)  &  $18.95_{-0.02} ^{0.02}$   & $18.95_{-0.03} ^{0.03}$  &  $18.92_{-0.04} ^{0.04}$ \\
           &  int. (ct/ks) &  $0.084_{-0.031} ^{2.71}$ & $0.088_{-0.04} ^{3.00}$ & $0.05_{-0.02} ^{29.00}$ \\
\hline
\end{tabular}
}
\end{center}
\end{table*}

\end{appendix}

\begin{thebibliography}{32}
\expandafter\ifx\csname natexlab\endcsname\relax\def\natexlab#1{#1}\fi

\bibitem[{{Babel} \& {Montmerle}(1997)}]{Babel1997}
{Babel}, J. \& {Montmerle}, T. 1997, \apjl, 485, L29+

\bibitem[{{Balona}(2013)}]{Balona2013}
{Balona}, L.~A. 2013, in Astronomical Society of the Pacific Conference Series,
  Vol. 479, Progress in Physics of the Sun and Stars: A New Era in Helio- and
  Asteroseismology, ed. H.~{Shibahashi} \& A.~E. {Lynas-Gray}, 385

\bibitem[{{Donati} {et~al.}(2006){Donati}, {Howarth}, {Jardine}, {Petit},
  {Catala}, {Landstreet}, {Bouret}, {Alecian}, {Barnes}, {Forveille},
  {Paletou}, \& {Manset}}]{Donati2006}
{Donati}, J.-F., {Howarth}, I.~D., {Jardine}, M.~M., {et~al.} 2006, \mnras,
  370, 629

\bibitem[{{Feldmeier} {et~al.}(1997{\natexlab{a}}){Feldmeier}, {Kudritzki},
  {Palsa}, {Pauldrach}, \& {Puls}}]{Feldmeier1997a}
{Feldmeier}, A., {Kudritzki}, R.-P., {Palsa}, R., {Pauldrach}, A.~W.~A., \&
  {Puls}, J. 1997{\natexlab{a}}, \aap, 320, 899

\bibitem[{{Feldmeier} {et~al.}(1997{\natexlab{b}}){Feldmeier}, {Puls}, \&
  {Pauldrach}}]{Feldmeier1997b}
{Feldmeier}, A., {Puls}, J., \& {Pauldrach}, A.~W.~A. 1997{\natexlab{b}}, \aap,
  322, 878

\bibitem[{{Fossati} {et~al.}(2015){Fossati}, {Castro}, {Sch{\"o}ller},
  {Hubrig}, {Langer}, {Morel}, {Briquet}, {Herrero}, {Przybilla}, {Sana},
  {Schneider}, {de Koter}, \& {BOB Collaboration}}]{Fossati2015}
{Fossati}, L., {Castro}, N., {Sch{\"o}ller}, M., {et~al.} 2015, \aap, 582, A45

\bibitem[{{Fossati} {et~al.}(2014){Fossati}, {Zwintz}, {Castro}, {Langer},
  {Lorenz}, {Schneider}, {Kuschnig}, {Matthews}, {Alecian}, {Wade}, {Barnes},
  \& {Thoul}}]{Fossati2014}
{Fossati}, L., {Zwintz}, K., {Castro}, N., {et~al.} 2014, \aap, 562, A143

\bibitem[{{Glebocki} \& {Gnacinski}(2005)}]{Glebocki2005}
{Glebocki}, R. \& {Gnacinski}, P. 2005, VizieR Online Data Catalog, 3244

\bibitem[{{Gray}(1992)}]{Gray88}
{Gray}, D.~F. 1992, The Observation and Analysis of Stellar Photospheres
  (Cambridge University Press)

\bibitem[{{G{\"u}del}(2004)}]{Guedel2004}
{G{\"u}del}, M. 2004, \aapr, 12, 71

\bibitem[{{Hubrig} {et~al.}(2017){Hubrig}, {Kholtygin}, {Ilyin},
  {Sch{\"o}ller}, \& {Jarvinen}}]{Hubrig2017}
{Hubrig}, S., {Kholtygin}, A.~F., {Ilyin}, I., {Sch{\"o}ller}, M., \&
  {Jarvinen}, S.~P. 2017, ArXiv e-prints [\eprint[arXiv]{1702.01063}]

\bibitem[{{Ignace} {et~al.}(2010){Ignace}, {Oskinova}, {Jardine}, {Cassinelli},
  {Cohen}, {Donati}, {Townsend}, \& {ud-Doula}}]{Ignace2010}
{Ignace}, R., {Oskinova}, L.~M., {Jardine}, M., {et~al.} 2010, \apj, 721, 1412

\bibitem[{{Malkov} {et~al.}(2012){Malkov}, {Tamazian}, {Docobo}, \&
  {Chulkov}}]{Malkov2012}
{Malkov}, O.~Y., {Tamazian}, V.~S., {Docobo}, J.~A., \& {Chulkov}, D.~A. 2012,
  \aap, 546, A69

\bibitem[{{Ness} {et~al.}(2002){Ness}, {Schmitt}, {Burwitz}, {Mewe}, {Raassen},
  {van der Meer}, {Predehl}, \& {Brinkman}}]{Ness2002}
{Ness}, J.-U., {Schmitt}, J.~H.~M.~M., {Burwitz}, V., {et~al.} 2002, \aap, 394,
  911

\bibitem[{{Nieva} \& {Przybilla}(2012)}]{Nieva2012}
{Nieva}, M.-F. \& {Przybilla}, N. 2012, \aap, 539, A143

\bibitem[{{Oskinova}(2016)}]{Oskinova2016}
{Oskinova}, L.~M. 2016, Advances in Space Research, 58, 739

\bibitem[{{Oskinova} {et~al.}(2014){Oskinova}, {Naz{\'e}}, {Todt},
  {Huenemoerder}, {Ignace}, {Hubrig}, \& {Hamann}}]{Oskinova2014}
{Oskinova}, L.~M., {Naz{\'e}}, Y., {Todt}, H., {et~al.} 2014, Nature
  Communications, 5, 4024

\bibitem[{{Owocki} {et~al.}(1998){Owocki}, {Cranmer}, \& {Gayley}}]{Owocki1998}
{Owocki}, S.~P., {Cranmer}, S.~R., \& {Gayley}, K.~G. 1998, \apss, 260, 149

\bibitem[{{Pillitteri} {et~al.}(2016){Pillitteri}, {Wolk}, {Chen}, \&
  {Goodman}}]{Pillitteri2016}
{Pillitteri}, I., {Wolk}, S.~J., {Chen}, H.~H., \& {Goodman}, A. 2016, \aap,
  592, A88

\bibitem[{{Pillitteri} {et~al.}(2014){Pillitteri}, {Wolk}, {Goodman}, \&
  {Sciortino}}]{Pillitteri2014c}
{Pillitteri}, I., {Wolk}, S.~J., {Goodman}, A., \& {Sciortino}, S. 2014, \aap,
  567, L4

\bibitem[{{Reale}(2007)}]{Reale2007}
{Reale}, F. 2007, \aap, 471, 271

\bibitem[{{Reale} {et~al.}(1997){Reale}, {Betta}, {Peres}, {Serio}, \&
  {McTiernan}}]{Reale1997}
{Reale}, F., {Betta}, R., {Peres}, G., {Serio}, S., \& {McTiernan}, J. 1997,
  \aap, 325, 782

\bibitem[{{Robrade} \& {Schmitt}(2011)}]{Robrade2011}
{Robrade}, J. \& {Schmitt}, J.~H.~M.~M. 2011, \aap, 531, A58

\bibitem[{{Sanz-Forcada} {et~al.}(2004){Sanz-Forcada}, {Franciosini}, \&
  {Pallavicini}}]{Sanz-Forcada2004}
{Sanz-Forcada}, J., {Franciosini}, E., \& {Pallavicini}, R. 2004, \aap, 421,
  715

\bibitem[{{Smith} \& {Gray}(1976)}]{Smith1976}
{Smith}, M.~A. \& {Gray}, D.~F. 1976, \pasp, 88, 809

\bibitem[{{Tonry} \& {Davis}(1979)}]{Tonry1979}
{Tonry}, J. \& {Davis}, M. 1979, \aj, 84, 1511

\bibitem[{{ud-Doula} \& {Owocki}(2002)}]{ud-Doula2002}
{ud-Doula}, A. \& {Owocki}, S.~P. 2002, \apj, 576, 413

\bibitem[{{Uesugi} \& {Fukuda}(1982)}]{Uesugi1982}
{Uesugi}, A. \& {Fukuda}, I. 1982, {Catalogue of stellar rotational velocities
  (revised)}

\bibitem[{{van Belle}(2012)}]{VanBelle2012}
{van Belle}, G.~T. 2012, \aapr, 20, 51

\bibitem[{{van Leeuwen}(2007)}]{VanLeeuwen2007}
{van Leeuwen}, F. 2007, \aap, 474, 653

\bibitem[{{Wade} {et~al.}(2014){Wade}, {Petit}, {Grunhut}, \&
  {Neiner}}]{Wade2014}
{Wade}, G.~A., {Petit}, V., {Grunhut}, J., \& {Neiner}, C. 2014, ArXiv e-prints
  [\eprint[arXiv]{1411.6165}]

\bibitem[Wade et al.(2017)]{Wade2016} Wade, G.~A., Shultz, M., Sikora, J., et al.\ 2017, \mnras, 465, 2517 
\end{thebibliography}
\end{document}